\documentclass[final]{IEEEtran}

\usepackage{amsthm,amssymb,graphicx,multirow,amsmath,color,amsfonts}
\usepackage[update,prepend]{epstopdf}
\usepackage[noadjust]{cite}
\usepackage[latin1]{inputenc}
\usepackage{tikz}
\usetikzlibrary{arrows,calc} 
\usepackage{bbm} 
\usepackage{pdfpages}
\usepackage{tabulary}
\usepackage{multirow}
\usepackage{comment}
\usepackage{setspace}
\usepackage{subfigure}
\usepackage{mathtools}

\def\nbr{{\mathbf{r}}}

\def\nb0{{\mathbf{0}}}
\def\nb1{{\mathbf{1}}}

\def\nbR{{\mathbf{R}}}

\def\nbbE{{\mathbb{E}}}

\def\sinc{{\rm sinc}}

\newtheorem{lemma}{Lemma}

\newtheorem{theorem}{Theorem}

\newtheorem{cor}{Corollary}

\newtheorem{remark}{Remark}

\allowdisplaybreaks 

\begin{document}
\graphicspath{{./Figures/}}
\title{Impact of UAV Wobbling on the Air-to-Ground Wireless Channel
}
\author{
Morteza Banagar, \textit{Student Member, IEEE}, Harpreet S. Dhillon, \textit{Senior Member, IEEE},\\ and Andreas F. Molisch, \textit{Fellow, IEEE}
\thanks{M. Banagar and H. S. Dhillon are with the Wireless@VT, Bradley Department of Electrical and Computer Engineering, Virginia Tech, Blacksburg, VA, USA (email: \{mbanagar, hdhillon\}@vt.edu), and A. F. Molisch is with the Wireless Devices and Systems Group, Ming Hsieh Department of Electrical and Computer Engineering, University of Southern California, Los Angeles, CA, USA (email: molisch@usc.edu). The support of the US NSF (Grants CNS-1923807 and CNS-1923601) is gratefully acknowledged.}
}

\maketitle
\vspace{-1.7cm}
\begin{abstract}
This paper studies the impact of unmanned aerial vehicle (UAV) wobbling on the coherence time of the wireless channel between UAVs and a ground user equipment (UE), using a Rician multi-path channel model. We consider two different scenarios for the number of UAVs: (i) \textit{single UAV scenario} (SUS), and (ii) \textit{multiple UAV scenario} (MUS). For each scenario, we model UAV wobbling by two random processes, i.e., the Wiener and sinusoidal processes, and characterize the channel autocorrelation function (ACF) which is then used to derive the coherence time of the channel. For the MUS, we further show that the UAV-UE channels for different UAVs are uncorrelated from each other. A key observation in this paper is that even for small UAV wobbling, the coherence time of the channel may degrade quickly, which may make it difficult to track the channel and establish a reliable communication link.
\end{abstract}


\section{Introduction} \label{sec:intro}

Unmanned aerial vehicles (UAVs) have recently gained significant attention in wireless communications due to their fast and cost-efficient deployment, mobility, and high probability of line-of-sight (LoS) \cite{J_Zeng_Wireless_2016}. UAVs can be deployed as a set of user equipment (UE) to perform various tasks, such as package delivery, monitoring, and surveillance, or as wireless relays or even base stations to complement the coverage and capacity of terrestrial cellular networks \cite{J_Mozaffari_Tutorial_2018}. Although the fast and easy deployment of UAVs is quite appealing for many applications, a significantly different operational regime motivates many fundamental questions. For instance, due to the lack of fixed and stable infrastructure and various environmental issues, such as bad weather conditions or wind gusts, UAVs may experience random wobbling (also termed fluctuations \cite{J_Dabiri_Analytical_2020} and jittering \cite{C_Xu_Robust_2018, J_Xu_Multiuser_2020, J_Wu_Energy_2020} in the relevant literature) while hovering at a specific location \cite{C_Li_Development_2017}. Although this wobbling is typically small (less than $10^\circ$ \cite{J_Ahmed_Flight_2010}), it could severely affect the quality of the wireless channel because of the large values of the carrier frequencies (any frequency of the order of or above $1$ GHz can be considered ``large" in the context of UAV wobbling). Quite remarkably, the impact of UAV wobbling on the properties of the air-to-ground wireless channel, such as its coherence time, has not been quantified yet in the literature, and is the main focus of this paper.

{\em Prior Art.} There has been a lot of recent interest in the analysis and design of UAV-assisted communication networks \cite{J_Simunek_UAV_2013, J_Chetlur_Downlink_2017, J_Amer_Mobility_2020, J_Morteza_Performance_2019, C_Morteza_3GPP_2019, C_Morteza_Fundamentals_2019, J_Morteza_Handover_2020}. However, only a handful of them considered the impact of random wobbling of UAVs on the performance of the UEs. In \cite{J_Dabiri_Analytical_2020}, the authors studied the transmitter-receiver antenna mismatch caused by the random wobbling of hovering UAVs in millimeter wave (mmWave) wireless communications. The problem of resource allocation in a drone cellular network when the UAVs are equipped with uniform linear antenna arrays and are also wobbling is studied in \cite{C_Xu_Robust_2018} and further extended to planar antenna arrays in \cite{J_Xu_Multiuser_2020}. In these works, the authors designed algorithms to minimize total power consumption in a multiple-input single-output (MISO) system by jointly optimizing the UAV trajectory and transmit beamforming vector. Taking UAV wobbling into account, the authors in \cite{J_Wu_Energy_2020} investigated a wiretap aerial system where the problem of secure and energy-efficient communication between a UAV and a ground UE is analyzed. In \cite{J_Liu_CoMP_2019}, the authors proposed a novel aerial network design of coordinated multipoint (CoMP) which benefits from both interference mitigation and UAV mobility \cite{J_Amer_Mobility_2020, J_Morteza_Performance_2019, C_Morteza_3GPP_2019, C_Morteza_Fundamentals_2019}. Specifically, they considered a Rician fading channel model where the LoS path has a random phase component due to the random UAV wobbling. Although these works address important problems related to UAV wobbling, its impact on the wireless communication channel still remains an open problem, which is the main focus of this paper.

{\em Contributions.} In this paper, we assume a Rician multi-path channel model and consider two scenarios for the number of UAVs: (i) single UAV scenario (SUS), where a single UAV communicates with the UE, and (ii) multiple UAV scenario (MUS), where multiple UAVs form a distributed multiple-input multiple-output (MIMO) transceiver to communicate with the ground UE. We then model the wobbling behavior of the UAVs by two random processes, i.e., the Wiener and sinusoidal processes. For the SUS, we rigorously characterize the channel autocorrelation function (ACF) for both random processes and determine the coherence time of the channel. We further derive the channel autocorrelation matrix in the MUS and demonstrate that the channels of different UAVs to the UE are uncorrelated from each other and the coherence time of these channels is the same as that of the SUS. Our analysis demonstrates that the choice of any realistic random process that captures the oscillatory nature of wobbling will result in a non-stationary received signal because of which the notion of channel coherence time needs to be defined carefully for this setting. A key design insight obtained from our analysis is that the coherence time of the channel is highly sensitive to UAV wobbling. Specifically, even for small wobbling of the UAV (w.r.t. the signal wavelength), we observe that the coherence time is not very large, which in turn makes the channel tracking and symbol detection difficult. To the best of our knowledge, this is the first work that characterizes the impact of UAV wobbling on the coherence time of the channel.

\section{System Model} \label{sec:SysMod}
\begin{figure}
	\centering
	\includegraphics[width=0.89\columnwidth]{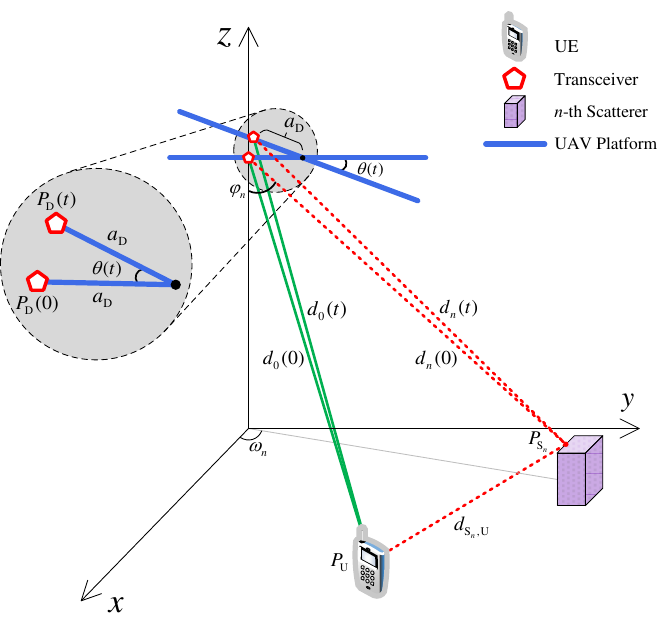}
	\vspace{-0.2cm}
	\caption{An illustration of the system model. The green solid lines represent the LoS links from the UAV transceiver to the UE at times $0$ and $t$ and the red dotted lines represent the MPCs from the scatterers.}
	\vspace{-0.2cm}
	\label{fig:SystemModel}
\end{figure}

We use the Cartesian coordinate system to represent the locations of the UAVs, the UE, and scatterers. For the number of UAVs, we consider two different scenarios: (i) SUS, where a single UAV is deployed at some arbitrary location to communicate with the UE, and (ii) MUS, where multiple UAVs are deployed at arbitrary locations to jointly communicate with the UE in a distributed-MIMO fashion. To isolate the effect of the UAV wobbling, the UE is assumed to be static. The UAVs are assumed to be rotary-winged drones that are hovering at their locations. As shown in Fig. \ref{fig:SystemModel} for the SUS, the UAV is equipped with a single antenna (transceiver) which is located under the UAV platform with an offset of $a_{\rm D}$ meters from its centroid. We assume that the ground is aligned with the $xy$-plane while the UAV platform is located in the $yz$-plane and is initially parallel to the $xy$-plane at height $h = z_{\rm D}$. The initial location of the transceiver is assumed to be $P_{\rm D}(0) = (0, 0, z_{\rm D})$ and the locations of the UE and the $n$-th scatterer are denoted by $P_{\rm U} = (x_{\rm U}, y_{\rm U}, 0)$ and $P_{{\rm S}_n} = (x_{{\rm S}_n}, y_{{\rm S}_n}, z_{{\rm S}_n})$, respectively (with the convention that $P_{{\rm S}_0} \equiv P_{\rm U}$). We represent the UAV-UE, UAV-scatterer, and scatterer-UE distances at time $t$ by $d_0(t)$, $d_n(t)$, and $d_{{\rm S}_n, {\rm U}}$, respectively. The angle-of-departure (AoD) from the UAV to the UE and from the UAV to the $n$-th scatterer (measured w.r.t. the $z$-axis) are denoted by $\varphi_0$ and $\varphi_n$, respectively. Furthermore, the angle between the $x$-axis and the line connecting the origin to the projection of $P_{{\rm S}_n}$ onto the ground is denoted by $\omega_n$.

We assume that the UAV may experience wobbling due to the lack of robust and fixed infrastructure, wind gusts, and the high vibration frequency of its propellers and rotors \cite{C_Li_Development_2017}, thus making it wobble. We model this wobbling by random processes and study their impact on the received signal at the UE. Note that because of this wobbling, the UAV platform may rotate in any of its three dimensions: roll, pitch, and yaw. In this paper, however, we only consider the pitch angle for simplicity as in \cite{C_Xu_Robust_2018, J_Xu_Multiuser_2020}. This pitch angle is denoted by $\theta(t)$ at time $t$.

In this paper, we assume a multi-path channel model where there is one LoS link between the UAV and the UE (the green solid lines in Fig. \ref{fig:SystemModel}) and $N$ multi-path components (MPCs) from scatterers (the red dotted lines for the $n$-th MPC in Fig. \ref{fig:SystemModel}). In the SUS, we represent the received signal $r(t)$ at time $t$ in the baseband (using the convention $d_{{\rm S}_0, {\rm U}} = 0$) as \cite{B_Molisch_Wireless_2011}
\begin{align}\label{eq:r_t}
r(t) = \sum_{n=0}^{N} \alpha_n {\rm e}^{-j \frac{2\pi}{\lambda} (d_n(t) + d_{{\rm S}_n, {\rm U}})},
\end{align}
where $\lambda = \frac{c}{f_{\rm c}}$ is the wavelength of the received signal, $c$ is the speed of light, $f_{\rm c}$ is the carrier frequency, and $\alpha_0$ and $\alpha_n$ are the amplitudes of the LoS link and the $n$-th MPC, respectively. Note that $\alpha_n$ and $d_n(t)$ may not be independent from each other in general. Similarly in the MUS, the received signal from the $i$-th UAV can be written as
\begin{align}\label{eq:r_t2}
r_i(t) = \sum_{n=0}^{N} \alpha_{i,n} {\rm e}^{-j \frac{2\pi}{\lambda} (d_{i,n}(t) + d_{{\rm S}_n, {\rm U}})},
\end{align}
where $\alpha_{i,n}$ and $d_{i,n}(t)$ are the amplitude and distance from the $i$-th UAV to the $n$-th scatterer. The AoD from the $i$-th UAV to the $n$-th scatterer is denoted by $\varphi_{i, n}$.

\begin{remark}\label{rem:0}
For notational ease, we assume that the antenna gain is constant within the range of interest for the angles of the MPCs. For a reasonably smooth amplitude antenna pattern $G(\varphi)$ and assuming small pitch angle, a nonuniform antenna pattern would only require multiplication of $\alpha_n$ with $G(\varphi_n)$.
\end{remark}

\begin{remark}\label{rem:1}
Due to the high carrier frequency of the received signal ($f_{\rm c}=1 \sim 6$ GHz), the wavelength will be in the order of centimeters ($\lambda = 5 \sim 30$ cm). Hence, a small variation in the UAV-UE distance may cause a large phase offset. Note that this effect is even more pronounced at mmWave frequencies.
\end{remark}

We now introduce the coherence time ($T_{\rm C}$) of the channel as the main metric of interest in this paper. Coherence time is defined as the time duration over which the impulse response of the channel is almost constant. Writing the channel ACF for the stationary received signal $r(t)$ as $R(\tau) = \nbbE[r(t)r^*(t+\tau)]$, a common way to define the channel coherence time is the first time instant when the normalized ACF ($R(\tau)/\max{R(\tau)}$) drops below a certain threshold $\gamma$ \cite{C_Va_Basic_2015}, i.e.,
\begin{align}\label{eq:T_C}
T_{\rm C} = \min\left\{\tau: \frac{R(\tau)}{\max{R(\tau)}} \leq \gamma\right\}.
\end{align}

\begin{remark}\label{rem:1.5}
When the received signal is non-stationary, the channel ACF becomes a function of both $t$ and $\tau$ and will be denoted as $R(t,t+\tau)$. We can define the coherence time in this case by first obtaining $T_C(t)$ from \eqref{eq:T_C} using $R(t,t+\tau)$ instead of $R(\tau)$ for all $t$ and then determining $T_C = \min_t T_C(t)$.
\end{remark}

\section{Channel Autocorrelation Function} \label{sec:Analysis}
In this section, we present a comprehensive analysis of the channel ACF for both the SUS and the MUS. We begin our analysis by deriving the relation between the pitch angles at time instants $t$ and $t+\tau$ and the difference between the UAV-scatterer (or UAV-UE) distances at these times in the following lemma.
\begin{lemma} \label{lem:DiffDist}
For a wobbling UAV, reasonably assuming $a_{\rm D} \ll d_n(t)$ and $\theta(t) \ll 1~{\rm rad}$, we have
\begin{align}\label{eq:dDn}
d_{{\rm D}, n}(t, t+\tau) :=&\, d_n(t+\tau) - d_n(t) \nonumber\\
\approx&\, a_{\rm D}\cos(\varphi_n)[\theta(t+\tau) - \theta(t)].
\end{align}
\end{lemma}
\begin{IEEEproof}
As shown in Fig. \ref{fig:SystemModel}, when the UAV wobbles, the location of the transceiver changes from $P_{\rm D}(0) = (0, 0, z_{\rm D})$ to $P_{\rm D}(t) = (0, a_{\rm D}(1-\cos(\theta(t))), z_{\rm D} + a_{\rm D}\sin(\theta(t)))$. We now write the equations for $d_n(0)$ and $d_n(t)$ as follows:
\begin{align*}
&d_n(0) = \sqrt{x_{{\rm S}_n}^2 + y_{{\rm S}_n}^2 + (z_{{\rm S}_n} - z_{\rm D})^2},\\
&d_n(t) = \scalebox{0.91}{$\sqrt{x_{{\rm S}_n}^2 \!\!+\! [y_{{\rm S}_n} \!\!-\! a_{\rm D}(1\!-\!\cos\theta(t))]^2 \!+\! [z_{{\rm S}_n} \!\!-\! z_{\rm D} \!-\! a_{\rm D}\!\sin\theta(t)]^2}$}\\
&\overset{(a)}{\approx} \!\!\scalebox{0.85}{$\sqrt{x_{{\rm S}_n}^2 \!\!\!+\! y_{{\rm S}_n}^2 \!\!\!+\! (z_{{\rm S}_n} \!\!\!-\! z_{\rm D})^2 \!- 2a_{\rm D}[y_{{\rm S}_n}\!(1\!-\!\cos\theta(t)) \!+\! (z_{{\rm S}_n} \!\!\!-\! z_{\rm D})\sin\theta(t)]}$}\\
&\overset{(b)}{\approx} d_n(0) - \frac{a_{\rm D}[y_{{\rm S}_n}(1-\cos\theta(t)) + (z_{{\rm S}_n} \!-z_{\rm D})\sin\theta(t)]}{d_n(0)},
\end{align*}
where in $(a)$ we used $a_{\rm D} \ll y_{{\rm S}_n}$ and in $(b)$ we used the approximation $\sqrt{1-\beta} \approx 1 - \frac{\beta}{2}$ for small $\beta$. Using $d_n(0) = (z_{\rm D}-z_{{\rm S}_n})/\cos(\varphi_n)$, we have $d_n(t) - d_n(0)$
\begin{align}\label{eq:NewApprox}
&\approx\frac{a_{\rm D}\cos(\varphi_n)}{z_{\rm D}-z_{{\rm S}_n}}\left[(z_{\rm D}-z_{{\rm S}_n})\sin\theta(t)-y_{{\rm S}_n}(1-\cos\theta(t))\right]\nonumber\\
&\overset{(a)}{=}a_{\rm D}\cos(\varphi_n)\sin\theta(t) - a_{\rm D} \sin(\omega_n)\sin(\varphi_n)(1-\cos\theta(t))\nonumber\\
&\approx a_{\rm D}\cos(\varphi_n)\theta(t) - a_{\rm D} \sin(\omega_n)\sin(\varphi_n)\frac{\theta^2(t)}{2}\nonumber\\
&\approx a_{\rm D}\cos(\varphi_n)\theta(t),
\end{align}
where in $(a)$ we used $y_{{\rm S}_n}/(z_{\rm D}-z_{{\rm S}_n}) = \tan(\varphi_n)\sin(\omega_n)$ and the last two steps result from the small-angle approximation of the pitch angle ($\theta(t) \ll 1~{\rm rad}$). The same approximation is also true for time instant $t+\tau$, i.e., $d_n(t+\tau) - d_n(0)\approx a_{\rm D}\cos(\varphi_n)\theta(t+\tau)$. Using this result and that in \eqref{eq:NewApprox}, the lemma is proved.
\end{IEEEproof}

\begin{cor}\label{cor:stationary}
Assuming $\theta(t)$ to be a random process with stationary increments, i.e., $\theta(t + \tau) - \theta(t)$ has the same distribution as $\theta(\tau)$, the result of Lemma \ref{lem:DiffDist} can be simplified as $d_{{\rm D}, n}(\tau) := d_{{\rm D}, n}(t, t + \tau) \approx a_{\rm D}\cos(\varphi_n)\theta(\tau)$.
\end{cor}

\subsection{ACF Analysis in the SUS}
We present the main result of this paper in the following theorem.
\begin{theorem} \label{thm:MainR}
The channel ACF for a wobbling UAV in a multi-path channel is given as
\begin{align}\label{eq:R_main}
R(t, t+\tau) = \sum_{n=0}^N \nbbE\left[|\alpha_n|^2{\rm e}^{j\frac{2\pi}{\lambda}a_{\rm D}\cos(\varphi_n)[\theta(t+\tau) - \theta(t)]}\right],
\end{align}
which is a function of both $t$ and $\tau$ (non-stationary ACF). However, if $\theta(t)$ has stationary increments, then the channel ACF becomes only a function of $\tau$ (stationary ACF):
\begin{align}\label{eq:R_main_cor}
R(\tau) &= \sum_{n=0}^N \nbbE\left[|\alpha_n|^2{\rm e}^{j\frac{2\pi}{\lambda}a_{\rm D}\cos(\varphi_n)\theta(\tau)}\right].
\end{align}
\end{theorem}
\begin{IEEEproof}
We can write the channel ACF using \eqref{eq:r_t} as
\begin{align}
R(t, t+\tau) &= \nbbE[r(t)r^*(t+\tau)]\nonumber\\
&\hspace{-1.5cm}=\nbbE\!\left[\sum_{m=0}^N\sum_{n=0}^N\scalebox{0.95}{$\alpha_m\alpha_n^*{\rm e}^{j\frac{2\pi}{\lambda}\left(d_n(t+\tau)-d_m(t)\right)}{\rm e}^{j\frac{2\pi}{\lambda}\left(d_{{\rm S}_n, {\rm U}} - d_{{\rm S}_m, {\rm U}}\right)}$}\right]\nonumber\\
&\hspace{-1.5cm}=\sum_{m=0}^N\!\sum_{\substack{n=0\\n\ne m}}^N \!\scalebox{0.95}{$\nbbE\!\left[\alpha_m\alpha_n^*{\rm e}^{j\frac{2\pi}{\lambda}\left(d_n(t+\tau)-d_m(t)\right)}\right]\! \mathbb{E}\!\left[{\rm e}^{j\frac{2\pi}{\lambda}\left(d_{{\rm S}_n, {\rm U}} - d_{{\rm S}_m, {\rm U}}\right)}\right]$}\nonumber\\
&\hspace{-1.5cm}+ \sum_{n=0}^N \nbbE\left[|\alpha_n|^2{\rm e}^{j\frac{2\pi}{\lambda}d_{{\rm D}, n}(t, t+\tau)}\right],
\end{align}
where the double summation in the last equality is zero since the random variable $\big[\frac{d_{{\rm S}_n, {\rm U}} - d_{{\rm S}_m, {\rm U}}}{\lambda} ~{\rm mod}~1\big]$ is uniformly distributed from $0$ to $1$ \cite[Lemma 4]{J_Dhillon_Backhaul_2015}. Hence, using Lemma \ref{lem:DiffDist} and Corollary \ref{cor:stationary}, we obtain \eqref{eq:R_main} and \eqref{eq:R_main_cor}, respectively.
\end{IEEEproof}
One can compute \eqref{eq:R_main} and \eqref{eq:R_main_cor} by first conditioning on $\varphi_n$, evaluating the resulting expectation, and then deconditioning for a given distribution of $\varphi_n$. While the result in \eqref{eq:R_main} holds for any angular power spectrum model, it would be instructive to simplify it for a specific model to obtain further insights. For that, we will use the well-accepted Laplacian model for the power of the $n$-th MPC \cite{J_Pedersen_Power_1997, B_Molisch_Wireless_2011}, which is given as $|\alpha_n|^2 = \frac{1}{2\sigma}{\rm e}^{-\frac{|\varphi_n - \varphi_0|}{\sigma}}$, $1\leq n \leq N$, where $\sigma$ is the scale parameter of the Laplacian model. In this paper, we use the Rician multi-path fading model with factor $K$ to capture the higher probability of LoS in aerial networks. The following corollary simplifies the result of Theorem \ref{thm:MainR} when $\theta(t)$ is a random process with stationary increments (so that ACF is stationary) and $|\alpha_n|^2$ follows a Laplacian model.
\begin{cor}\label{cor:SUS1}
Assuming the Laplacian angular power spectrum with $|\alpha_0|^2 = K \sum_{n=1}^N |\alpha_n|^2 = K \sum_{n=1}^N \frac{1}{2\sigma}{\rm e}^{-\frac{|\varphi_n - \varphi_0|}{\sigma}}$ and that $\theta(t)$ has stationary increments, the channel ACF can be simplified as
\begin{align}\label{eq:R_LoS}
R(t, t + \tau) = \,&R(\tau) = \nbbE\left[|\alpha_0|^2{\rm e}^{j\frac{2\pi}{\lambda}a_{\rm D}\cos(\varphi_0)\theta(\tau)}\right] +\nonumber\\
&\sum_{n=1}^N \nbbE\left[\frac{1}{2\sigma}{\rm e}^{-\frac{|\varphi_n - \varphi_0|}{\sigma}}{\rm e}^{j\frac{2\pi}{\lambda}a_{\rm D}\cos(\varphi_n)\theta(\tau)}\right].
\end{align}
\end{cor}

As mentioned in the previous section, due to the UAV wobbling, we model the variations in the pitch angle by random processes. Assuming the wobbling of the pitch angle has stationary increments, i.e., the resulting ACF is a function of the time difference $\tau$ only, this wobbling imposes an \emph{effective Doppler shift} of $f_{\rm D} = \frac{a_{\rm D}\cos(\varphi_0)\theta(\tau)}{\lambda \tau}$ on the wireless channel. In this paper, we use two different random processes for this purpose: (i) the Wiener process, and (ii) the sinusoidal process. Note that one can also consider more complex random processes, such as the periodic Brownian bridge or constrained Wiener processes to analyze the coherence time of the channel, however, the coherence time analysis may not even be tractable in the resulting non-stationary settings for these processes.

\begin{remark}\label{rem:2}
\emph {(No wobbling).} In an ideal setting where the UAV platform is ``completely" stable without any wobbling or angular deviations, we have $\theta(t) = 0$ which results in a constant value for $R(t, t+\tau)$ for all $t$ and $\tau$. As expected, the coherence time will be infinity in this ideal case.
\end{remark}

\subsubsection{Wiener Process}
The fundamental properties of a Wiener process $W(t)$ can be summarized as follows: (i) $W(0) = 0$, (ii) $W(t)$ has independent, stationary, and Gaussian increments, (iii) $W(t)$ is continuous in $t$. Assuming $\theta(t)$ to be a Wiener process, one can show that it is scale-invariant and its probability density function (pdf) follows a Gaussian distribution with mean zero and variance $bt$, where $b~{\rm rad}^2/{\rm s}$ is a proportionality constant which can also be used as a tuning parameter. Hence, following Theorem \ref{thm:MainR}, we have
\begin{align}\label{eq:R_Wiener}
R(t, t + \tau) &= R(\tau) = \sum_{n=0}^N \nbbE\left[|\alpha_n|^2\nbbE\left[{\rm e}^{j\frac{2\pi}{\lambda}a_{\rm D}\cos(\varphi_n)W(\tau)}\right]\right]\nonumber\\
&= \sum_{n=0}^N \nbbE\left[|\alpha_n|^2{\rm e}^{-\left( \frac{2\pi^2}{\lambda^2}a_{\rm D}^2 \cos^2(\varphi_n) \right)b\tau}\right],
\end{align}
where the last equality results from the characteristic function (cf) of a Gaussian random variable. Note that since the Wiener process has stationary increments, we were able to use \eqref{eq:R_main_cor} in Theorem \ref{thm:MainR} and the ACF is only a function of the time difference $\tau$. This result shows that the channel ACF becomes an exponentially decaying function of $\tau$ when the pitch angle is modeled as a Wiener process, which severely affects the coherence time of the channel. 


\subsubsection{Sinusoidal Process}
In this case, we assume that the pitch angle is given by $\theta(t) = A \sin(2\pi F t)$, where $A$ and $F$ are independent random variables representing the amplitude and the frequency of the pitch angle variations, respectively. In this paper, we assume $A \sim U[-\theta_{\rm m}, \theta_{\rm m})$ and $F \sim p_F(f)$, where $\theta_{\rm m}$ is the maximum pitch angle and $p_F(.)$ is some given pdf. Note that the sinusoidal random process does not have the stationary increment property. Hence, we obtain the channel ACF using Theorem \ref{thm:MainR} as $R(t, t+\tau) =$
\begin{align}\label{eq:R_Sin}
&= \sum_{n=0}^N \nbbE\left[|\alpha_n|^2{\rm e}^{j\frac{2\pi}{\lambda}a_{\rm D}\cos(\varphi_n)A [\sin(2\pi F (t+\tau))-\sin(2\pi F t)]}\right]\nonumber\\
&= \sum_{n=0}^N \nbbE\bigg[|\alpha_n|^2\int_{-\infty}^\infty\sinc\Big(\frac{2}{\lambda}a_{\rm D}\cos(\varphi_n)\theta_{\rm m}\,\times\nonumber\\
&\hspace{1.4cm}\big[\sin(2\pi F (t+\tau)) -\sin(2\pi F t)\big]\Big)p_F(f) {\rm d}f\bigg],
\end{align}
where $\sinc(x) = \frac{\sin(\pi x)}{\pi x}$ and in the last equality, we used the cf of the uniform random variable and then took the expectation w.r.t. $F$. As we will see in Section \ref{sec:Sim}, the special case of $t = 0$ gives the lowest coherence time among different values of $t$. The channel ACF for $t = 0$ is further simplified as $R(0, \tau) = $
\begin{align}\label{eq:R_Sin_t0}
\sum_{n=0}^N \nbbE\!\left[|\alpha_n|^2\!\!\int_{-\infty}^\infty\!\!\!\!\sinc\left(\!\frac{2}{\lambda}a_{\rm D}\cos(\varphi_n)\theta_{\rm m}\sin(2\pi f \tau)\!\right)p_F(f) {\rm d}f\right]\!.
\end{align}
Note that the channel ACF is not a periodic function of $\tau$ due to the random frequency of the pitch angle. However, if we assume a constant $F$, then the channel ACF will be periodic in $\tau$ even with a random pitch angle amplitude $A$.

Using \eqref{eq:T_C}, \eqref{eq:R_Wiener}, and \eqref{eq:R_Sin}, one can now derive the coherence time of the channel for each random process. Explicitly for the Wiener process in an LoS channel ($N=0$) with an arbitrary \textit{non-random} AoD, the channel ACF is given as
\begin{align}\label{eq:TcWin0}
R(\tau) &= \nbbE\left[|\alpha_0|^2{\rm e}^{-\left( \frac{2\pi^2}{\lambda^2}a_{\rm D}^2 \cos^2(\varphi_0) \right)b\tau}\right]\nonumber\\
&= |\alpha_0|^2{\rm e}^{-\left( \frac{2\pi^2}{\lambda^2}a_{\rm D}^2 \cos^2(\varphi_0) \right)b\tau}.
\end{align}
Now since \eqref{eq:TcWin0} is a monotonically decreasing function of $\tau$ with its peak at $\tau = 0$, solving $\frac{R(\tau)}{\max{R(\tau)}} = \gamma$ gives the coherence time of the channel in closed-form as
\begin{align}\label{eq:TcWin}
T_{\rm C} = \frac{\lambda ^ 2}{2 b\pi^2 a_{\rm D}^2 \cos^2(\varphi_0)} \log\left(\frac{1}{\gamma}\right).
\end{align}
For the sinusoidal process, on the other hand, the closed-form solution is not available and we need to numerically solve \eqref{eq:R_Sin_t0} to obtain the coherence time of the channel.

\begin{remark}\label{rem:3}
\emph {(Impact of the Rician $K$-factor).} Using the assumptions and result of Corollary \ref{cor:SUS1}, we can write the channel ACF as a function of $K$ as
\begin{align*}
R(\tau; K) = \,&K \sum_{n=1}^N \nbbE\left[\frac{1}{2\sigma}{\rm e}^{-\frac{|\varphi_n - \varphi_0|}{\sigma}}{\rm e}^{j\frac{2\pi}{\lambda}a_{\rm D}\cos(\varphi_0)\theta(\tau)}\right] \,+\\
&\sum_{n=1}^N \nbbE\left[\frac{1}{2\sigma}{\rm e}^{-\frac{|\varphi_n - \varphi_0|}{\sigma}}{\rm e}^{j\frac{2\pi}{\lambda}a_{\rm D}\cos(\varphi_n)\theta(\tau)}\right].
\end{align*}
Assuming $\theta(\tau)$ follows the Wiener process, we have
\begin{align}\label{eq:rem3}
\frac{R(\tau; K)}{R(0; K)} = \,&\frac{K}{K+1} {\rm e}^{-\left( \frac{2\pi^2}{\lambda^2}a_{\rm D}^2 \cos^2(\varphi_0) \right)b\tau} \,+\nonumber\\
& \frac{\sum\limits_{n=1}^N \nbbE\left[{\rm e}^{-\frac{|\varphi_n - \varphi_0|}{\sigma}} {\rm e}^{-\left( \frac{2\pi^2}{\lambda^2}a_{\rm D}^2 \cos^2(\varphi_n) \right)b\tau} \right]}{(K+1)\sum\limits_{n=1}^N \nbbE\left[{\rm e}^{-\frac{|\varphi_n - \varphi_0|}{\sigma}} \right]}.
\end{align}
Taking the derivative of \eqref{eq:rem3} w.r.t. $K$, we end up with an expression which could be either negative or positive depending on the value of $\varphi_0$. Hence, the coherence time of the channel is a function of both $K$ and $\varphi_0$, which is neither increasing nor decreasing w.r.t. $K$.
\end{remark}

\begin{figure*}[!t]
	\centering
	\begin{minipage}{0.66\columnwidth}
		\centering
		\includegraphics[width=1\textwidth]{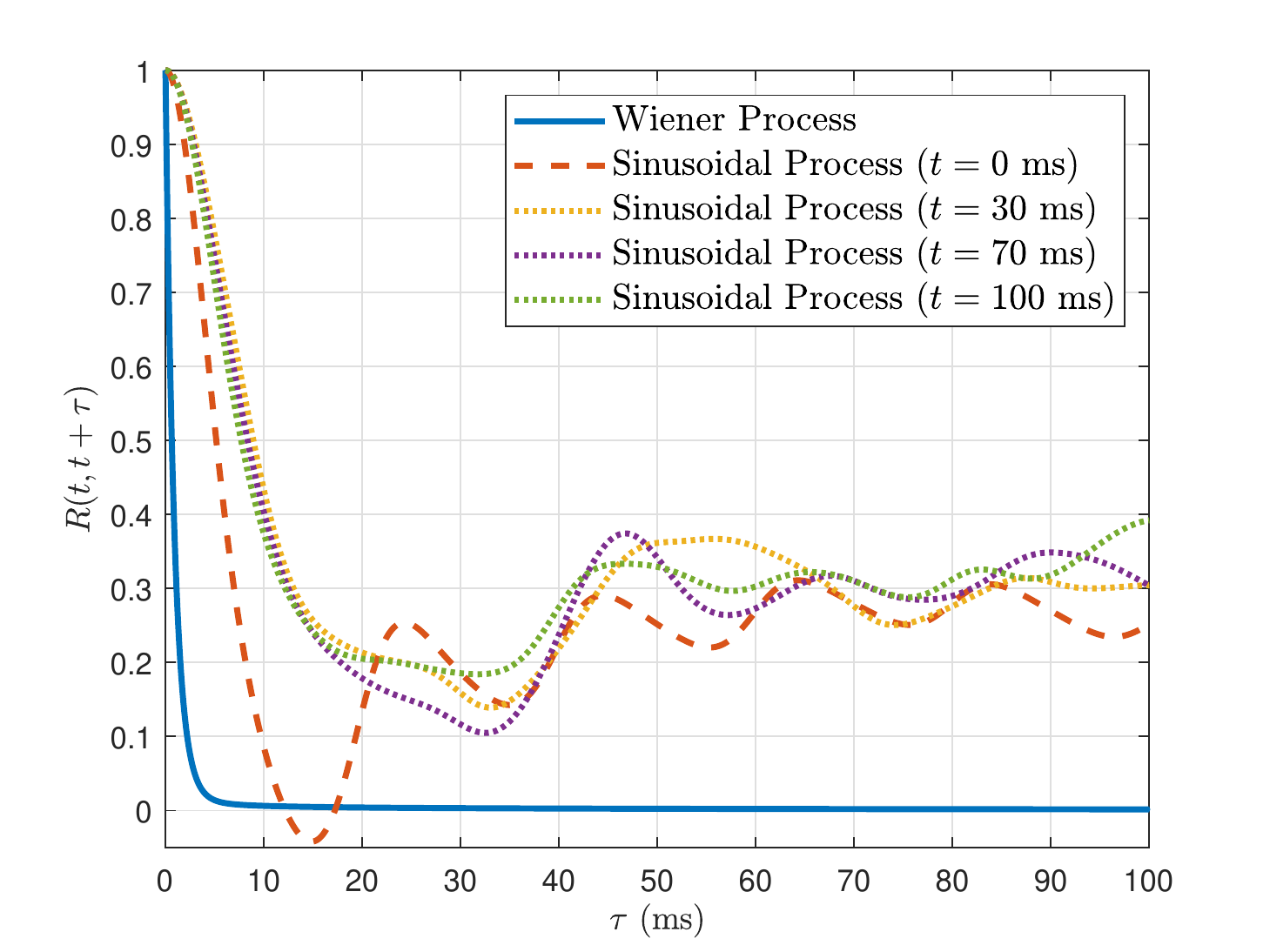}
		\vspace{-0.7cm}
		\caption{The channel ACF for different random processes. The parameters are $f_{\rm c} = 6$ GHz and $\theta_{\rm m} = 5^\circ$.}
		\vspace{-0.3cm}
		\label{fig:sim1}
	\end{minipage}\hfill
	\begin{minipage}{0.66\columnwidth}
		\centering
		\includegraphics[width=1\textwidth]{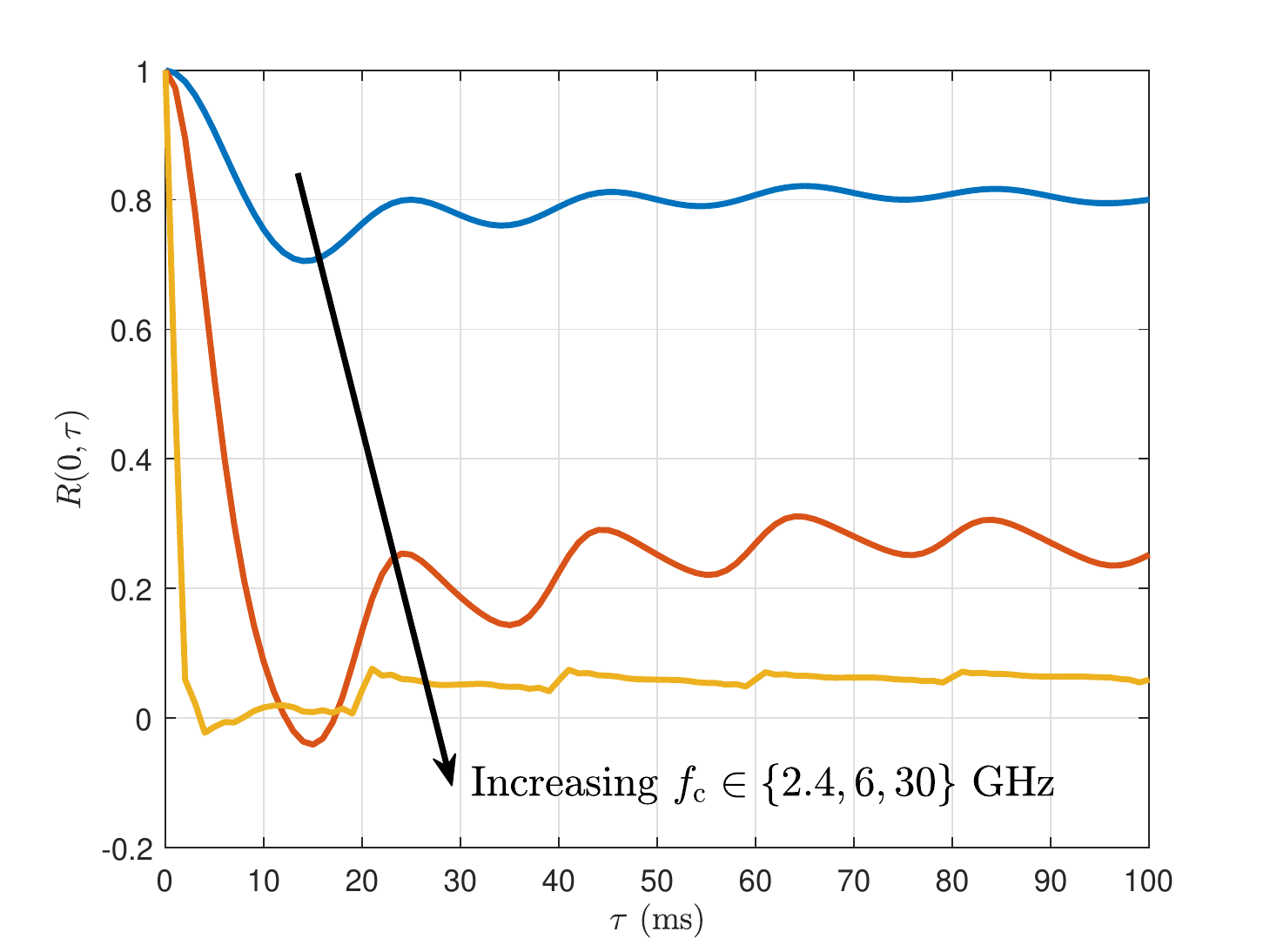}
		\vspace{-0.7cm}
		\caption{The channel ACF when $\theta(t)$ follows the sinusoidal process with varying carrier frequencies and $\theta_{\rm m} = 5^\circ$.}
		\vspace{-0.3cm}
		\label{fig:sim2}
	\end{minipage}\hfill
	\begin{minipage}{0.66\columnwidth}
		\centering
		\includegraphics[width=1\textwidth]{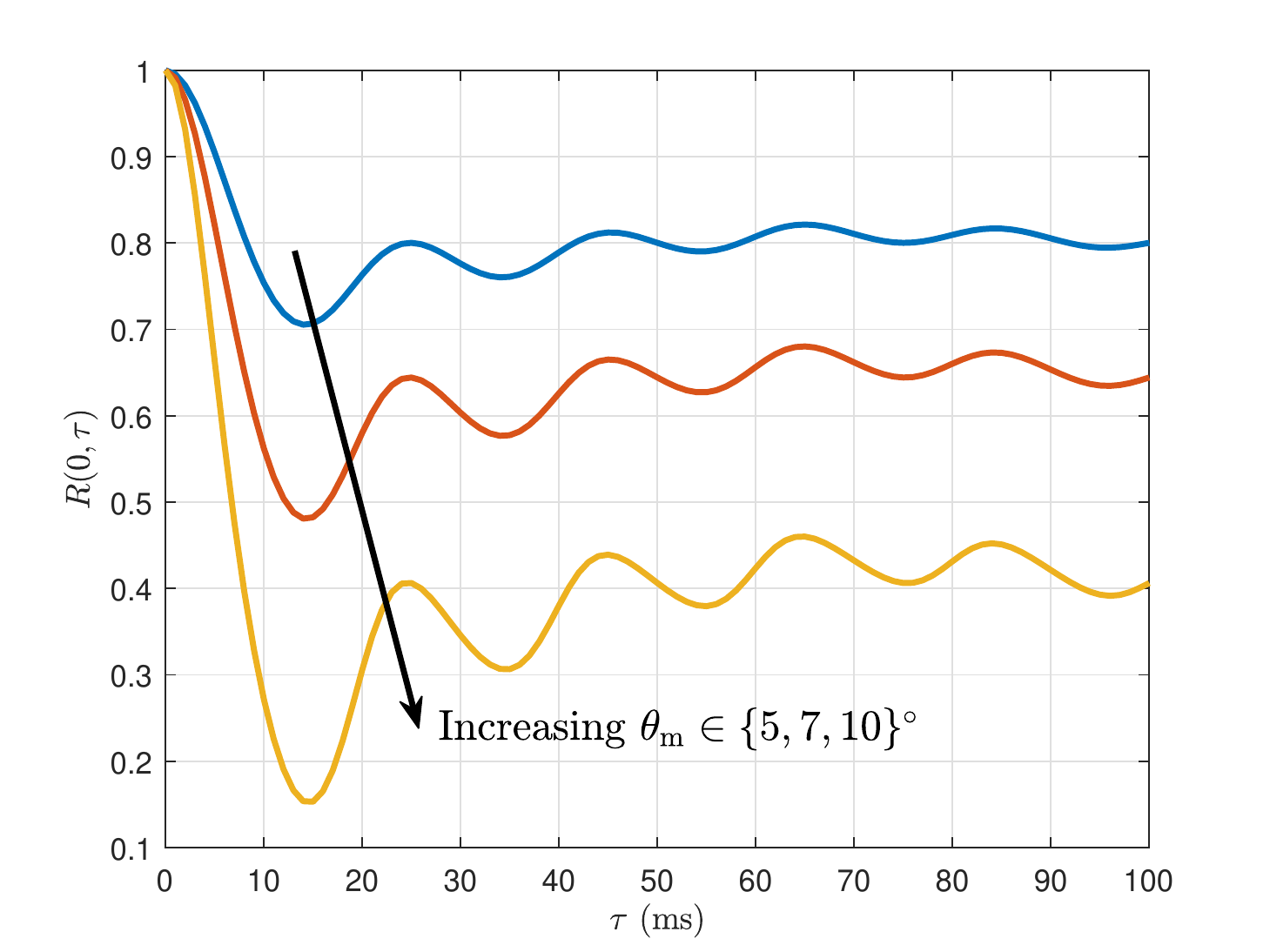}
		\vspace{-0.7cm}
		\caption{The channel ACF when $\theta(t)$ follows the sinusoidal process with varying maximum pitch angles and $f_{\rm c} = 2.4$ GHz.}
		\vspace{-0.3cm}
		\label{fig:sim3}
	\end{minipage}
\end{figure*}
\subsection{ACF Analysis in the MUS}
In this model, we assume that $M$ UAVs form a distributed MIMO transceiver to jointly communicate with the UE. The channel autocorrelation matrix can be written as
\begin{align*}
\nbR(t, t + \tau) = [R_{ik}(t, t + \tau)]_{1\leq i,k\leq M} = \nbbE\left[\nbr(t)\nbr^*(t+\tau)\right],
\end{align*}
where $\nbr(t) = [r_1(t), r_2(t), \dots, r_M(t)]^{\rm T}$ is the received signal vector, and $R_{ik}(t, t + \tau)$ is the $(i, k)$-th element of $\nbR(t, t + \tau)$.
\begin{theorem} \label{thm:MainR2}
The channel autocorrelation matrix for $M$ wobbling UAVs that form a distributed MIMO transceiver in an environment with one LoS link for each UAV and $N$ MPCs can be written as $\nbR(t, t + \tau) = $
\begin{align}\label{eq:R_main2}
{\rm diag}\left\{ \sum_{n=0}^N \nbbE\left[|\alpha_{i,n}|^2{\rm e}^{j\frac{2\pi}{\lambda}a_{\rm D}\cos(\varphi_{i, n})[\theta(t+\tau) - \theta(t)]}\right] \right\}.
\end{align}
\end{theorem}
\begin{IEEEproof}
For the diagonal elements of $\nbR(t, t + \tau)$, Theorem \ref{thm:MainR} is directly applied and we have
\begin{align*}
R_{ii}(t, t + \tau) &= \nbbE\left[r_i(t)r_i^*(t+\tau)\right]\\
&= \sum_{n=0}^N \nbbE\left[|\alpha_{i,n}|^2{\rm e}^{j\frac{2\pi}{\lambda}a_{\rm D}\cos(\varphi_{i, n})[\theta(t+\tau) - \theta(t)]}\right].
\end{align*}
On the other hand, for the off-diagonal elements, we can write
\begin{align*}
&R_{ik}(t, t + \tau) = \nbbE\left[r_i(t)r_k^*(t+\tau)\right]\\
&= \nbbE\!\left[\sum_{m=0}^N\sum_{n=0}^N \scalebox{0.9}{$\alpha_{i, m}\alpha_{k,n}^*{\rm e}^{j\!\frac{2\pi}{\lambda}\left(d_{k, n}(t+\tau)-d_{i, m}(t)\right)}{\rm e}^{j\!\frac{2\pi}{\lambda}\left(d_{{\rm S}_n, {\rm U}} - d_{{\rm S}_m, {\rm U}}\right)}$}\right]\nonumber\\
&= \!\sum_{n=0}^N \scalebox{0.9}{$\nbbE\!\left[\alpha_{i, n}\alpha_{k,n}^*{\rm e}^{j\!\frac{2\pi}{\lambda}\left(d_{k, n}(t+\tau)-d_{k, n}(t)\right)}\right]\!\nbbE\!\left[{\rm e}^{j\!\frac{2\pi}{\lambda}\left(d_{k, n}(t)-d_{i, n}(t)\right)}\right]$} +  \nonumber\\
&\sum_{m=0}^N\!\sum_{\substack{n=0\\n\ne m}}^N\!\!\scalebox{0.9}{$\nbbE\!\left[\alpha_{i, m}\alpha_{k,n}^*{\rm e}^{j\!\frac{2\pi}{\lambda}\left(d_{k, n}(t+\tau)-d_{i, m}(t)\right)}\right]\!\nbbE\!\left[{\rm e}^{j\!\frac{2\pi}{\lambda}\left(d_{{\rm S}_n, {\rm U}} - d_{{\rm S}_m, {\rm U}}\right)}\right]$},
\end{align*}
where in the first summation of the last equality we used the fact that the pitch angle wobbling in the location of UAVs is independent of the distances between the UAVs and the scatterers (or the UE). Hence, since the random variable $\big[\frac{d_{k, n}(t)-d_{i, n}(t)}{\lambda} ~{\rm mod}~1\big]$ is uniformly distributed from $0$ to $1$ \cite[Lemma 4]{J_Dhillon_Backhaul_2015}, we conclude that this summation is zero. The second summation is also zero using a similar reasoning as in the proof of Theorem \ref{thm:MainR}.
\end{IEEEproof}
\begin{cor}\label{cor:MUS1}
Assuming the Laplacian angular power spectrum with Rician fading model, and that $\theta(t)$ has stationary increments, the channel autocorrelation matrix is simplified as
\begin{align}\label{eq:R_LoS21}
&\!\!\!\nbR(t, t \!+\! \tau) \!=\! \nbR(\tau) \!=\! {\rm diag}\left\{\nbbE\left[|\alpha_{i, 0}|^2{\rm e}^{j\frac{2\pi}{\lambda}a_{\rm D}\cos(\varphi_{i, 0})\theta(\tau)}\right]\right\} + \nonumber\\
&\sum_{n=1}^N {\rm diag}\left\{\nbbE\left[\frac{1}{2\sigma}{\rm e}^{-\frac{|\varphi_{i, n} - \varphi_{i, 0}|}{\sigma}}{\rm e}^{j\frac{2\pi}{\lambda}a_{\rm D}\cos(\varphi_{i, n})\theta(\tau)}\right]\right\}.
\end{align}
\end{cor}
Similar to the SUS, one can model the random process $\theta(t)$ using the Wiener or sinusoidal processes to obtain the channel autocorrelation matrix. The fundamental observation in the MUS is that when multiple UAVs are hovering at some locations to communicate with the UE in a distributed-MIMO fashion, then the channels will be uncorrelated from each other. Using \eqref{eq:T_C}, \eqref{eq:R_main}, and \eqref{eq:R_main2}, we conclude that the coherence time of the channel in the MUS is the same as that of the SUS.

\vspace{-0.1cm}
\section{Simulation Results} \label{sec:Sim}
In this section, we present numerical results to demonstrate the impact of UAV pitch wobbling on the coherence time of the channel. We assume that a single rotary-winged UAV hovers at some arbitrary location and wobbles based on either the Wiener or sinusoidal processes. For the number of scatterers, we assume $N = 20$ and $N = 10$ in the sub-$6$ GHz and mmWave frequencies, respectively \cite{B_Molisch_Wireless_2011}. Following Corollary \ref{cor:SUS1}, we assume a Laplacian angular power spectrum with $K = 11.5$ \cite[Table 1]{C_Goddemeier_Investigation_2015}, $\varphi_0 = 20^\circ$ and $\sigma = 1$. For the Wiener process, we assume the proportionality constant is $b = 1 ~{\rm rad}^2/{\rm s}$ and for the sinusoidal process, we assume that the amplitude and frequency of the pitch angle both follow the uniform distribution, i.e., $A \sim U[-\theta_{\rm m}, \theta_{\rm m})$ and $F \sim U[5, 25)$ Hz. Other parameters are $a_{\rm D} = 40$ cm, $\theta_{\rm m} = \{5, 7, 10\}^\circ$ (the maximum pitch angle of $10^\circ$ is selected based on \cite{J_Ahmed_Flight_2010}), and $f_{\rm c} = \{2.4, 6, 30\}$ GHz (equivalently, $\lambda = \{12.5, 5, 1\}$ cm). Note that the derived values for the channel coherence time should be treated as useful ballpark figures, since they are dependent on the physical characteristics of the UAVs and may vary from one UAV to another.


In Fig. \ref{fig:sim1}, we show the impact of UAV pitch angle wobbling on the channel ACF for both random processes where $f_{\rm c} = 6$ GHz and $\theta_{\rm m} = 5^\circ$. Note that since the channel ACF is non-stationary in the sinusoidal process, we plotted the ACF at different values of $t$ to understand its behavior. As seen in this figure, the sinusoidal process represents its lowest coherence time at $t=0$ s. Thus, in the remainder of this section, we use this time instant to calculate the coherence time of the channel for the sinusoidal process. Assuming a normalized threshold of $\gamma = 0.5$, the coherence time of the channel is $642~\mu {\rm s}$ and $5.18~{\rm ms}$ for the Wiener and sinusoidal processes, respectively. Note that the observed behavior for the Wiener process is intuitive since the variations in the pitch rotation angle can grow without bounds, and thus, the channel decorrelates with itself rapidly, yielding a very low coherence time. Nevertheless, it is interesting to note that the coherence time for the sinusoidal model is not too high either. Hence, channel tracking and phase estimation for the proper symbol detection become very difficult in both models \cite{C_Zhu_Tracking_2015}. Comparing these two random processes, we observe that the Wiener process is tractable and has stationary increments, but suffers from unbounded and non-smooth angular variations. On the other hand, the sinusoidal process is reasonably realistic and has bounded and smoother angular variations at the cost of being relatively less tractable and not having stationary increments. In fact, any reasonably realistic model that captures the oscillatory behavior of the wobbling motion of UAVs is unlikely to possess stationary increments.

In Figs. \ref{fig:sim2} and \ref{fig:sim3}, we examine the sinusoidal model more carefully and demonstrate the impact of the carrier frequency and the maximum pitch angle on the channel ACF and the coherence time. From a physical standpoint, since the impulse response of the channel depends on the ratio of the transmitter-receiver distance $d$ and the signal wavelength $\lambda$, a higher $d/\lambda$ translates to a higher channel variation, which in turn results in a lower coherence time. Thus, increasing $\theta_{\rm m}$ (higher $d$) or increasing $f_{\rm c}$ (lower $\lambda$), gives a lower channel coherence time. Note that increasing the UAV antenna offset $a_{\rm D}$ would have the same impact on the coherence time of the channel. Furthermore, increasing the range of the UAV pitch wobbling frequency $F$ makes the channel to decorrelate with itself more rapidly, which also yields a lower coherence time. Note that the limiting value of the stationary ACF ($R(\tau)$ as $\tau \to \infty$) also decreases with increasing $\theta_{\rm m}$ or $f_{\rm c}$. Assuming a normalized threshold of $\gamma = 0.5$ and $\theta_{\rm m} = 5^\circ$, we observe from Fig. \ref{fig:sim2} that $T_{\rm C} = \{\infty, 5.18, 0.97\}~{\rm ms}$ for $f_{\rm c} = \{2.4, 6, 30\}$ GHz, respectively. As it can be seen in this figure, with the physical parameters \cite{C_Goddemeier_Investigation_2015, J_Ahmed_Flight_2010} mentioned in the beginning of this section, the coherence time of the channel will be in the order of microseconds for mmWave frequencies. Hence, channel tracking and, in turn, communication is very challenging at the mmWave frequencies. From Fig. \ref{fig:sim3}, we have $T_{\rm C} = \{\infty, 12.26, 6.74\}~{\rm ms}$ for $\theta_{\rm m} = \{5, 7, 10\}^\circ$, respectively, when $f_{\rm c} = 2.4$ GHz. Consequently, when working in higher frequencies or in inclement weather, we require UAVs with robuster stabilizers that guarantee very low angular deviations.



\vspace{-0.2cm}
\section{Conclusion} \label{sec:Conclusion}
In this paper, we provided a rigorous mathematical analysis for the coherence time of the channel when UAVs experience random pitch wobbling. Assuming a Rician multi-path channel model, we considered two different scenarios for the number of UAVs communicating with the UE, i.e., SUS and MUS, and modeled the UAV pitch wobbling by random processes. For both the SUS and MUS, we formulated the channel ACF and derived the coherence time of the channel. Specifically in the MUS, we showed that the channels between the UAVs and the UE are uncorrelated from each other and the channel autocorrelation matrix is only determined by its diagonal elements. Our analysis demonstrated that even for small UAV pitch wobbling, the coherence time of the channel could be severely affected, thus making channel tracking and symbol detection difficult. A meaningful extension of this work is to study the impact of UAV wobbling on the fundamental characteristics of the channel when the UAVs are \textit{mobile}. Another direction for future work is to derive the coherence time of the channel in a centralized MIMO scenario, i.e., having an antenna array instead of a single antenna in the UAV structure \cite{C_Xu_Robust_2018, J_Xu_Multiuser_2020}.


\bibliographystyle{IEEEtran}
\bibliography{../../AllReferences}
\end{document}